\documentclass[10pt,aps,prd,twocolumn,floats,floatfix,showpacs,superscriptaddress,nofootinbib,longbibliography]{revtex4-2}
\usepackage[utf8]{inputenc}               		

\usepackage{calligra}
\usepackage{graphicx,mathtools,amssymb,amsmath,amsthm,amsfonts,epsfig,epsf,soul}
\usepackage[outdir=./]{epstopdf}
\usepackage[dvipsnames]{xcolor}
\usepackage[linktocpage]{hyperref}
\hypersetup{
    colorlinks=true,
    linkcolor=blue,
    filecolor=magenta,      
    urlcolor=BrickRed,
    citecolor=BrickRed,
}
\hypersetup{pdfstartview=}	
\usepackage{tensor}	
\usepackage{csquotes}
\usepackage{mathrsfs}
\usepackage{float}
\usepackage[normalem]{ulem}

\usepackage{subfig}
\usepackage{graphicx}
\usepackage{amsmath}
\usepackage{autobreak}

\begin{document}

\allowdisplaybreaks

\title{Exotic compact objects and light bosonic fields}

\author{\textbf{Farid Thaalba}}
\affiliation{Nottingham Centre of Gravity \& School of Mathematical Sciences, University of Nottingham,
University Park, Nottingham NG7 2RD, United Kingdom}
\author{\textbf{Giulia Ventagli}}
\affiliation{Nottingham Centre of Gravity \& School of Mathematical Sciences, University of Nottingham,
University Park, Nottingham NG7 2RD, United Kingdom}
\affiliation{CEICO, Institute of Physics of the Czech Academy of Sciences, Na Slovance 2, 182 21 Praha 8, Czechia}
\author{\textbf{Thomas P.~Sotiriou}}
\affiliation{Nottingham Centre of Gravity \& School of Mathematical Sciences, University of Nottingham,
University Park, Nottingham NG7 2RD, United Kingdom}
\affiliation{School of Physics and Astronomy, University of Nottingham,
University Park, Nottingham NG7 2RD, United Kingdom}

\begin{abstract}
In this note, we discuss the effect of light, non-gauge, bosonic degrees of freedom on the exterior spacetime of an exotic compact object. We show that such fields generically introduce large deviations from black hole spacetimes of General Relativity near and outside the surfaces of ultra-compact exotic objects unless one assumes they totally decouple from the standard model or new heavy fields. Hence, using black hole spacetimes of General Relativity to model ultra-compact exotic objects and their perturbations relies implicitly on this assumption or on the absence of such fields. 
\end{abstract}

\maketitle

The landmark detection by the LIGO and Virgo collaboration \cite{LIGOScientific:2016aoc} of gravitational waves (GWs) emitted by coalescing compact objects, and the explosion in the number of recorded events of GWs that followed \cite{LIGOScientific:2018mvr,LIGOScientific:2020ibl,LIGOScientific:2021usb,LIGOScientific:2021djp} has opened up the prospect of testing the nature and dynamics of the most compact objects in the universe. This exploration will be furthered by future observations \cite{Barack:2018yly,Sathyaprakash:2019yqt,Barausse:2020rsu,Kalogera:2021bya,LISA:2022kgy}, which will probe some of the most violent phenomena in the universe in an effort to identify deviations from general relativity (GR) in the form of non-linear interactions between gravity and new degrees of freedom (d.o.f), and possible quantum effects that might affect the structure of black holes (BHs).

Classically, the formation of black holes as the endpoints of gravitational collapse is a well-understood process. Already by the late 1930s, there were strong indications that black holes should exist in nature, such as the  Chandrasekhar limit for the mass of a white dwarf~\cite{Chandrasekhar:1935zz} and the Oppenheimer-Volkoff limit for the mass of a new star~\cite{Oppenheimer:1939ne}, and can form from collapse, such as the Oppenheimer and Snyder collapse of pressureless homogeneous dust model~\cite{PhysRev.56.455}.
It was later shown by Penrose and Hawking that a spacetime which meets a set of appropriate causality and energy conditions must be either timelike or null incomplete \cite{Penrose:1964wq,Hawking:1971vc}. This led Penrose to propose his cosmic censorship conjecture~\cite{Penrose:1969pc} where such singularities are hidden inside a black hole. 

The existence of BHs has been confirmed by several observations, amongst these are that of Sgr $A\ast$, a supermassive black hole at the centre of the Milky Way~\cite{10.1093/mnras/291.1.219,Ghez_1998}, as well as the images of M$87\ast$ and Sgr $A\ast$ obtained by the Event Horizon Telescope~\cite{EventHorizonTelescope:2022wkp, Vagnozzi:2022moj}. Additionally, gravitational wave observations from the LIGO-Virgo-KAGRA collaboration are fully consistent with the description of BH binaries \cite{KAGRA:2021vkt}.
Nevertheless, with observations now promising to probe compact objects more precisely than ever, a lot of effort is being put into testing with higher and higher accuracy if the objects we expect to be black holes are indeed black holes. 

The term Exotic Compact Objects (ECOs) is used to refer collectively to compact objects that can resemble black holes (see \cite{Cardoso:2019rvt} for a review). Boson stars are an example of an ECO for which there is a complete classical description \cite{Seidel:1993zk,Liebling:2012fv}. They exist in theories where an axion or an axion-like particle is minimally coupled to gravity where the new degree of freedom gives rise to a self-gravitating soliton. 
But for the most part, ECOs are conjectured to exist based on theoretical arguments that suggest quantum effects, most probably nonlocal, might become large near the horizon. This then leads to a significant deviation from the Kerr spacetime near the horizon and in the interior. A characteristic example is that of fuzzballs \cite{Mathur:2005zp,Mayerson:2020tpn}, arising from a combination of string theory considerations and attempts to solve the black hole information paradox~\cite{Hawking:1975vcx,Mathur:2008nj}.

Since in most cases an explicit description of the ECO, as well as equations that govern its dynamics, are not available, a common approach to test the black hole paradigm with gravitational waves is to use a phenomenological description of an ECO. In this approach, which will be our focus here, one typically assumes that all corrections introduced by the new unknown physics are confined to the interior of the ECO and simply vanish in the exterior. This leads to the equations 
\begin{align}
    G_{\mu\nu} + T_{\mu\nu}^\text{int} = 0,
    \label{eq1}
\end{align}
where $T_{\mu\nu}^\text{int}$ describes the interior of the ECO and includes the corrections introduced by the new unknown physics, while it vanishes in the exterior. If one further assumes spherical symmetry, the exterior is described by the Schwarzschild metric, due to Birkhoff's theorem and the main difference between ECOs and BHs is taken to be that the surface of an ECO is not completely absorbent but it possesses some reflective properties. 
The boundary conditions on the surface of an ECO are different than those on the horizon of the black hole, which implies that quasinormal modes (QNMs) are modified even when the exterior is identical. A distinctive feature of certain ECOs is the appearance of late-time echoes due to the formation of quasi-bound states between the photon sphere and the surface of the ECO~\cite{Kojima:1995nc,Ferrari:2000sr}.
For a more detailed discussion of ECOs and their phenomenology, we refer the reader to \cite{Cardoso:2019rvt}.

Assuming that the new physics is entirely confined inside the star, in analogue to compact stars that indeed have a surface, is rather reasonable when one postulates that only new fermionic fields are involved. Beyond spherical symmetry, the exterior does not have to be close to the Kerr metric and the internal structure would in principle be imprinted on the multipolar structure of the exterior. This complicates the theory-agnostic framework for more realistic configurations, but it is perhaps also an opportunity to probe the interior. Here we want to draw attention to a different caveat: that light, non-gauge, bosonic degrees of freedom will generically endow ECOs with new charges that can significantly affect the exterior. Hence, the agnostic approach described above and widely used in the literature {\em implicitly} assumes that they are either entirely absent or that there is some mechanism that suppresses these charges.
 
For concreteness, we start by considering a  massless scalar field in a Schwarzschild background with no backreaction governed by 
\begin{align}
    \label{eq:scalar_eqn}
    \Box \phi = 0, \quad \Box = \nabla_{\mu} \nabla^{\mu}, 
\end{align}
where the only solution of the scalar equation is $\phi = \text{constant}$. This is due to requiring regularity of the scalar field at the horizon, as integrating the previous equation once yields
\begin{align}
    \label{eq:scalar_derivative}
    \partial_r \phi = \frac{C}{r(r-2M)},
\end{align}
where $M$ is the ADM mass of the Schwarzschild BH, $C$ is an integration constant. Therefore, $C$ must be zero for regularity, leaving only constant $\phi$ solutions. This holds more generally due to no-hair theorems \cite{Chase:1970omy,Hawking:1972qk,Bekenstein:1995un,Sotiriou:2011dz,Hui:2012qt,Sotiriou:2015pka,Herdeiro:2015waa} and indeed one can see the regularity of the scalar on the event horizon as the essential assumption of these no-hair theorems.  

Let us now entertain the thought that there is no horizon, but instead, there is a surface at some radius larger than $2M$. Inside that surface, Eq.~\eqref{eq:scalar_eqn} on a Schwarzschild background ceases to be a good effective description. Then there is no obvious reason for which $C$ has to vanish. As we now moved the surface closer to $r=2M$ the gradient of $\phi$, and hence the backreaction on the spacetime, grows, which contradicts the assumption that Schwarzschild would be a good description of the spacetime all the way to the surface. The deviation can become large even for very small values of $C$ if the surface is pushed close to $2M$.

This simple example clearly illustrates the issue we want to highlight, which is however not specific to massless scalars. A light scalar, where light in this context means that the inverse mass is larger than the size of the horizon, would behave the same way near the horizon. But also, the same issue would arise for higher spin non-gauge fields. Consider a Proca field for example. Using the Stueckelberg trick (see Ref.~\cite{Ruegg:2003ps} for a detailed review), one can restore gauge symmetry. The additional degree of freedom with respect to those of a gauge field is then made explicit. The Lagrangian then takes the form~\cite{Stueckelberg:1938hvi}
\begin{align}
    \mathcal{L} = -\frac{1}{4} F_{\mu\nu}F^{\mu\nu}-\frac{1}{2}(m A_\mu + \partial_\mu \phi)(m A^\mu + \partial^\mu \phi),
\end{align}
where $F_{\mu\nu} = \partial_\mu A_\nu - \partial_\nu A_\mu$. In the decoupling limit $m\to 0$, the scalar d.o.f. decouples from the vector field and satisfies Eq.~\eqref{eq:scalar_eqn}.

It is also worth emphasizing that the diverging behaviour of the field on the horizon is a rather generic feature and not specific to the exact form of Eq.~\eqref{eq:scalar_eqn}. That is to say, it does originate from the behaviour of the Laplacian on the horizon, which will be generically present if the theory is to satisfy second-order differential equations, but it persists if other terms are present. For example, in scalar-tensor theories that evade no-hair theorems, regulating the divergence fixes the value of the scalar charge \cite{Kanti:1995vq,Sotiriou:2013qea,Sotiriou:2014pfa}. The same pattern holds for new charges in other theories that exhibit black holes hairs, such as Einstein-aether theory and Ho\v{r}ava-Lifshitz gravity (in which case the diverging behaviour might be on the horizon of some effective metric)~\cite{Barausse:2011pu,Barausse:2013nwa}. 

We have illustrated qualitatively why the absence of a horizon implies that, generically, the light field would be non-trivial and have significant back-reaction near and outside the location where the horizon would otherwise be.  Let us now turn our attention to BH mimickers i.e., ECOs with a surface that is very close to the would-be horizon, and try to quantify how large this back-reaction would be and how it affects the metric.  We assume that in the exterior of the surface of the ECO the following equations hold
\begin{align}
    G_{\mu \nu} &= -\frac{1}{2}g_{\mu \nu}(\nabla\phi)^2 + \nabla_{\mu}\phi\nabla_{\nu}\phi \eqqcolon T_{\mu \nu}^{\phi}, \label{eq:GR}\\
    \Box \phi &= 0.
    \label{eq:scalar}
\end{align}
This is consistent with the usual assumptions that all new physics is confined in the interior, except we have allowed for the new light scalar.
Furthermore, for simplicity, we assume staticity and spherical symmetry which lead to the ansatz 
\begin{align}
    \mathrm{d} s^2&=-A(r)\mathrm{d}t^2+B(r)^{-1}\mathrm{d}r^2+r^2\mathrm{d}\Omega^2, \\ 
    \phi &= \phi(r).
\end{align}
Following a similar approach to Ref.~\cite{Sotiriou:2014pfa}, we parametrize the deviations from the Schwarzschild metric by performing a perturbative expansion in some bookkeeping parameter $c$
\begin{align}
\label{eq:perturbative_expansion}
    A(r)=&\left(1-\frac{2m}{r}\right)\left(1+\sum_{n=1}^{\infty}A_n(r)c^n\right)^2,\\
    B(r)=&\left(1-\frac{2m}{r}\right)\left(1+\sum_{n=1}^{\infty}B_n(r)c^n\right)^{-2},\\
    \phi(r)=&\;\sum_{n=0}^\infty \phi_n c^n, 
\end{align}
where $m$ is a parameter that reduces to the ADM mass of the spacetime in the case of $c=0$. Solving Einstein's field equations and the scalar equation to second order in $c$, we find
\begin{align}
    \begin{split}
    -g_{tt}(r) &= 1-\frac{2M}{r}\\
    &+\frac{Q^2 \left[(M-r) \log \left(1-\frac{2 M}{r}\right)-2 M\right]}{4 M^2 r}, 
    \end{split}\\
    g_{rr}(r) &= \frac{r}{r-2 M}+\frac{Q^2 r \log \left(1-\frac{2 M}{r}\right)}{4 M (r-2 M)^2}, \\ 
    \phi(r) &= \phi _0+\frac{Q \log \left(\frac{r}{r-2 M}\right)}{2 M}.
\end{align}
where $M$ is the ADM mass of the ECO's spacetime, $Q$ is the scalar charge, that is the $1/r$ coefficient that appears in the expansion of the scalar field at infinity as a power series in $1/r$ i.e., $\phi(r\rightarrow\infty) = \phi_{\infty} + \frac{Q}{r}+\mathcal{O}\left(1/r^2\right)$. For the perturbative equations and the relation between $M, Q, m$, and $c$ see appendix \ref{Equations}. It is worth noting that eqs.~\eqref{eq:GR} and \eqref{eq:scalar} are known to admit the Janis-Newman-Winicour metric as an exact solution~\cite{Janis:1968zz,Virbhadra:1997ie}. Our perturbative solution agrees to second order in $Q$ with a small charge expansion of this exact solution. We restrict ourselves to perturbative treatment, as it is more transparent and easy to follow.

Let us now study the behaviour of the metric functions and the scalar field at the surface of an ECO, $r_{\text{surface}}$, with a radius arbitrarily close to the would-be horizon. We then set $r=r_{\text{surface}}=2 M (1+\epsilon)$ for $\epsilon\ll1$. To order $\mathcal{O}(c^2)$, we have
\begin{align}
   -g_{tt}(r=r_{\text{surface}}) & = \frac{Q^2 (-\log (\epsilon ))-2 Q^2}{8 M^2}+\mathcal{O}\left(\epsilon\right), \\
   g_{rr}(r=r_{\text{surface}}) &= \frac{Q^2 \log (\epsilon )}{8 M^2 \epsilon ^2}+\mathcal{O}\left(\frac{1}{\epsilon }\right), \\
   \phi(r=r_{\text{surface}}) &= \phi _0+\frac{Q \log \left(\frac{1}{\epsilon }\right)}{2 M}+\mathcal{O}\left(\epsilon \right).
\end{align}
The behavior of the Kretschmann invariant $\mathcal{K}$ and the trace of the energy-momentum tensor $T^{\phi}$ near the would-be horizon is 
\begin{align}
    \begin{split}
    \mathcal{K}(r=&r_{\text{surface}})= \frac{Q^2}{16 M^6 \epsilon }\\
    &+\frac{3 \left(4 M^2-Q^2 \log \left(\frac{1}{\epsilon}\right)-2 Q^2\right)}{16 M^6}+\mathcal{O}\left(\epsilon\right), 
    \end{split}\\ 
    T^{\phi}(r=&r_{\text{surface}}) = -\frac{Q^2}{16 M^4 \epsilon }+\frac{3 Q^2}{16 M^4}+\mathcal{O}\left(\epsilon\right).
\end{align}

Hence,  for the Schwarzschild metric to adequately describe the geometry outside of an ECO, we need to have $Q/M \ll \epsilon$.\par 
If we consider for illustrative purposes BH mimickers with a very small closeness parameter of order the Planck length i.e., $\epsilon \sim l_{P}$, then the scalar charge per unit mass must at least be of the same order for the corrections to the metric to be ``small". If instead we only ask for the values of $T^{\phi}$ or $\mathcal{K}$ to be bounded and remain under control, then we would need for $Q/M \sim \sqrt{\epsilon}$ which still requires significant fine-tuning. Furthermore, without fine-tuning the charge per unit mass, the ECO would experience very large curvature in the exterior as given by the Kretschmann scalar, and to support such curvatures the interior would need to exhibit immense stress to prevent the collapse into a BH. Additionally, we mention, that if we consider only radial perturbations, then the potential appearing in the usual Schrodinger-like equation \cite{Blazquez-Salcedo:2018jnn, Antoniou:2022agj} behaves near the would-be horizon as $V(r=r_{\text{surface}}) \sim -Q^2/(M^4\epsilon) + \mathcal{O}(1)$ which might indicate radial instability unless, again, $Q/M$ is fine-tuned. These results hold equally well for scalars with a Compton wavelength much larger than the size of the object. 

If Eq.~\eqref{eq:scalar} were to remain unmodified in the interior, $Q$ would have to vanish by regularity at the centre. Indeed, a no-hair theorem for stars and for shift-symmetric scalars has been established in Ref.~\cite{Lehebel:2017fag}. However, setting $Q$ to zero by invoking such a theorem, or arguing that it is natural for it to be extremely small, requires making assumptions about the physics in the interior. In particular, it requires assuming that if new light fields exist, they are essentially decoupled from the standard model or new heavy fields in the interior of the ECO, which is by no means generic. 

To conclude, in this note, we have highlighted a possible subtlety of the effect light bosonic fields have on the exterior spacetime of an ECO. 
In the literature, it is often assumed that vacuum GR governs the gravitational interaction in the exterior of an ECO and that, invoking spherical symmetry, for $r>r_{\text{surface}}$ the geometry is adequately described by the Schwarzschild solution. This assumption seems to rely heavily on the idea that the physical processes inside the object and/or presumable quantum effects that prevent the formation of a horizon do not affect the geometry of the exterior. What we have demonstrated here is that this expectation is not true for ultra-compact ECOs if the physics governing their interior includes light bosonic fields that do not totally decouple from the standard model or new heavy fields. 

Our arguments do not rely on the nature or specific interactions of the light bosonic fields, nor do they assume that they are crucial for the prevention of collapse and the formation of the horizon - they merely have to be present. This is because if they exist, their configurations tend to diverge near the would-be horizon unless the ECO carries zero charges associated with these fields.
The fact that we have not yet detected new light fields in non-gravitational experiments does suggest that, if they exist, they couple weakly to the standard model at low curvatures. However, to assume that the ECO carries zero charge translates to a much stronger assumption about the coupling of these light fields, the standard model fields, and new heavy fields in the interior of the ECO. A possible workaround would be to devise a screening mechanism where the nonlinearities in the system conceal the scalar field all the way to the surface of the smallest ECO. Doing so would allow for a non-perturbative restoration of GR outside the ECO.

For simplicity, we have used spherical symmetry in our analysis. Relaxing this assumption and assuming that the exterior is described by the Kerr metric does not remove the issue we point out regarding light fields. For other vacuum spacetimes of general relativity the properties of the interior are encoded in the multipolar structure of the exterior. Axisymmetric spacetimes are a characteristic example, where one has a plethora of solutions with different properties already in general relativity \cite{Weyl:1917gp,papapetrou,Zipoy:1966btu,Ernst:1967wx,Stephani:2003tm}. Hence, even if Eq.~\eqref{eq1} would act as a good approximation for the exterior, it would be very tenuous to suggest that any particular solution of these equations can generically act as an adequate approximation for the spacetime of an ECO near its surface.

Finally, we have focused on stationary solutions, but it is clear that a modification of the exterior spacetime of the ECO would also affect both their QNM spectrum and gravitational wave signals from systems that contain such ECOs more broadly.  Attention has focused mostly on QNMs as probes of ECOs, which implicitly assumes that when ECOs merge the product is itself an ECO, rather than a black hole.  An interesting question, motivated further by our results, is how deviations manifesting in the late inspiral compare in size with deviation in the QNMs.

\section*{Acknowledgement}
We would like to thank Pedro Fernandes, Andrea Maselli and Silke Weinfurtner for useful feedback on earlier versions of this manuscript. G.V. acknowledges support from the Czech Academy of Sciences under grant number LQ100102101.
T.P.S. acknowledges partial support from
the STFC Consolidated Grants no.~ST/X000672/1 and
no.~ST/V005596/1. 
\appendix
\section{Perturbative EOM}
\label{Equations}
Here, we present the perturbed equations of motion to second order in $c$. To first order, the equations read:
\begin{align}
    0 & = B_1-(2 m-r) B_1', \\
    0 & = B_1+(2 m-r) A_1', \\
    0 & =(m-r) B_1'+(m+r) A_1'+r (r-2 m) A_1'', \\
    0 & = 2 (r-m) \phi _1'+r (r-2 m) \phi _1'',
\end{align}
while to second-order the equations are 
\begin{align}
    \begin{split}
    0 =& 4 r (r-2 m)^2 \left[(2 m-r) B_2'-B_2\right]\\ 
    &+c_3{}^2 (r-2 m)-6 c_1{}^2 r  
    \end{split}\\
    \begin{split}
    0 =& 4 r (r-2 m)^2 \left[(2 m-r) A_2'+B_2\right]\\ 
    &+c_3{}^2 (r-2 m)-2 c_1{}^2 r 
    \end{split}\\
    \begin{split}
    0 =& 2 r (r-2 m)^2 \big[ (m+r) A_2'+r (r-2 m) A_2''\\
    &+(m-r) B_2'\big]+c_3{}^2 (r-2m)-2 c_1{}^2 r 
    \end{split}\\
    \begin{split}
    0 =& (r-2 m)^2 \left[2 (r-m) \phi _2'+r (r-2 m) \phi _2''\right]\\
    &-2 c_1 c_3.
    \end{split}
\end{align}
Imposing the boundary conditions that $A_{1,2}, B_{1,2}, \phi_{1,2}$ vanish at spatial infinity and solving the equations yields
\begin{align}
    A_1 &= \frac{c_1}{r-2 m}\\
    B_1 &= \frac{c_1}{2 m-r} \\
    \phi _1 &=-\frac{c_3 \log \left(\frac{r}{r-2 m}\right)}{2 m}, \\
    \begin{split}
    A_2 &= \Bigg\{c_3{}^2 \left(2 m^2-3 m r+r^2\right) \log \left(\frac{r}{r-2 m}\right)\\
    &-2 m \big[-4 c_4 m (r-2 m)+2 c_1{}^2 m-2 c_3{}^2 m\\
    &+c_3{}^2 r\big]\bigg\}/\big[8 m^2 (r-2 m)^2\big], 
    \end{split}\\
    \begin{split}
    B_2 &= \bigg[4 m \left(4 c_4 m-2 c_4 r+3 c_1{}^2\right)+c_3{}^2 (2 m-r)\\
    &\times\log \left(\frac{r}{r-2m}\right)\bigg]/\big[8 m (r-2 m)^2\big],
   \end{split}\\
    \phi _2 &= \frac{\frac{2 c_1 c_3 m}{r-2 m}-(c_6 m+c_1 c_3) \log \left(\frac{r}{r-2 m}\right)}{2 m^2}.
\end{align}
where $c_{1}, c_{3}, c_{4}, c_{6}$ are integration constants. 
Performing asymptotic expansion at spatial infinity of the metric functions and the scalar field to $\mathcal{O}(c^2)$ we have 
\begin{widetext}
\begin{align}
    -g_{tt}(r \rightarrow \infty)|_{\mathcal{O}(c^2)} &= 1+\frac{-2 m+2 c c_1 +2 c^2 c_4}{r}+\frac{c^2 c_3{}^2 m}{6
   r^3}+\mathcal{O}\left(\frac{1}{r}\right)^4, \\ 
  g_{rr}(r \rightarrow \infty)|_{\mathcal{O}(c^2)} &= 1+\frac{2m-2c c_1-2c^2 c_4}{r}+\frac{\frac{1}{2} c^2 \left(-16 c_4 m+8 c_1{}^2-c_3{}^2\right)-8 c c_1 m+4m^2}{r^2}+& \nonumber \\
  &\frac{\frac{1}{2} c^2 \left(-48 c_4 m^2+48 c_1{}^2 m-5 c_3{}^2 m\right)-24 c c_1 m^2+8 m^3}{r^3}+\mathcal{O}\left(\frac{1}{r}\right)^4,\\
   \phi(r \rightarrow \infty)|_{\mathcal{O}(c^2)} &= \frac{-c_3 c-c_6 c^2}{r}+\frac{-m c_3 c+(c_1 c_3-m c_6)
   c^2}{r^2}+&\\
   &\frac{-\frac{4}{3} \left(m^2 c_3\right) c-\frac{4}{3} \left[m
   (m c_6-2 c_1 c_3)\right]c^2}{r^3}+\mathcal{O}\left(\frac{1}{r}\right)^4,
\end{align}
\end{widetext}
then to second order in $c$, the ADM mass $M$ and the scalar charge $Q$ are given by 
\begin{align}
    M &= m-cc_{1}-c^2c_{4}, \\ 
    Q &= -cc_{3}-c^2c_{6}.
\end{align}

\vspace{5mm}

\bibliography{biblio.bib}
\end{document}